\documentstyle[12pt]{article}

\input epsf.sty
\newdimen\psfigsize
\def\psfigure#1 #2 #3 #4 #5{
    \begin{figure}[tbp]
    \vbox{
    \null\hskip#2\epsfxsize=#1 \epsfbox[0 0 4096 4096]{#4}
    \vskip 10truept
    \caption {#5 \label{#3}}
    \vskip 0.1truein plus0.2truein}
    \end{figure}
}
\def\pspagefigure#1 #2 #3 #4 #5{
    \begin{figure}[p]
    \vbox{
    \null\hskip#2\epsfxsize=#1 \epsfbox[0 0 4096 4096]{#4}
    \vskip 10truept
    \caption {#5 \label{#3}}
    \vskip 0.1truein plus0.2truein}
    \end{figure}
}
\def\psoddfigure#1 #2 #3 #4 #5 #6{
    \begin{figure}[tbhp]
    \vbox{
    \null\hskip#3\epsfxsize=#1 \epsfbox[0 0 4096 4096]{#5}
    \vskip -#1 \vskip #2 \vskip 10truept
    \vskip 10truept
    \caption {#6 \label{#4}}
    \vskip 0.1truein plus0.2truein}
    \end{figure}
}
\def\figurespace#1 #2 #3 #4 {
    \begin{figure}[tbhp]
    \vbox{
    \psfigsize=#1truein
    \vskip \psfigsize
    \vskip 10truept
    \caption {#4 \label{#3}}
    \vskip 0.1truein plus0.2truein}
    \end{figure}
}
\def\gnufigure#1 #2 #3 #4 #5 #6{
    \begin{figure}[tbhp]
    \vbox{
    \null\hskip#3\epsfxsize=#1 \epsfbox{#5}
    \vskip -#1 \vskip #2 \vskip 10truept
    \vskip 10truept
    \hbox{\null\hskip 1.0in \parbox[t]{4.5in}{ \caption {#6 \label{#4}} } }
    \vskip 0.1truein plus0.2truein}
    \end{figure}
}

\def\pbp{\bar\psi\psi}

\def\psibar {\bar \psi}

\def\etal{{\it et al.}}

\def\khat{\hat k}

\def\xvec{{\vec x}}

\newcommand{\kc}{\kappa_c}

\def\LP{\left(}		
\def\RP{\right)}	

\def\rightpartial{{\overrightarrow\partial}}
\def\leftpartial{{\overleftarrow\partial}}
\def\diffpartial{\buildrel\leftrightarrow\over\partial}

\def\BE{\begin{equation}}
\def\EE{\end{equation}}
\def\BEA{\begin{eqnarray}}
\def\EEA{\end{eqnarray}}
\def\EL{\nonumber\\}
  
\newcommand{\gbeta}{6/g^2}

\newcommand{\ONEMP}{1^{-+}}
\newcommand{\ONEMPTWO}{1^{-+}_2}
\newcommand{\ZEROPM}{0^{+-}}
\newcommand{\ZEROMM}{0^{--}}
\newcommand{\QQQQ}{Q^4}
 
\begin{document}

\begin{titlepage}
\baselineskip=16pt
\rightline{\bf hep-lat/9707008}
\baselineskip=20pt plus 1pt
\vskip 1.5cm

\centerline{\Large \bf Exotic mesons in quenched lattice QCD}
\bigskip
\centerline{\bf Claude~Bernard and James~E.~Hetrick }
\centerline{\it
Department of Physics, Washington University, St.~Louis, MO 63130, USA
}
\centerline{\bf Thomas~A.~DeGrand and Matthew~Wingate }
\centerline{\it
Physics Department, University of Colorado, Boulder, CO 80309, USA
}
\centerline{\bf Carleton~DeTar and Craig~McNeile }
\centerline{\it
Physics Department, University of Utah, Salt Lake City, UT 84112, USA
}
\centerline{\it and}
\centerline{\it
Zentrum f\"ur Interdisziplin\"are Forschung, Universit\"at Bielefeld, Bielefeld,
  Germany}
\centerline{\bf Steven~Gottlieb}
\centerline{\it
Department of Physics, Indiana University, Bloomington, IN 47405, USA
}
\centerline{\bf Urs~M.~Heller }
\centerline{\it
SCRI, Florida State University, Tallahassee, FL 32306-4130, USA
}
\centerline{\bf Kari~Rummukainen }
\centerline{\it Universit\"at Bielefeld, Fakult\"at f\"ur Physik, Postfach
100131, D-33501 Bielefeld, Germany
}
\centerline{\bf Bob~Sugar }
\centerline{\it
Department of Physics, University of California, Santa Barbara, CA 93106, USA
}
\centerline{\bf Doug~Toussaint }
\centerline{\it
Department of Physics, University of Arizona, Tucson, AZ 85721, USA}
\centerline{\it and}
\centerline{\it
Center for Computational Physics, University of Tsukuba, Ibaraki 305, Japan }

\narrower
Since gluons in QCD are interacting fundamental constituents
just as quarks are, we expect that in addition to mesons made from a
quark and an antiquark, there should also be glueballs and hybrids
(bound states of quarks, antiquarks and gluons).
In general, these states would mix strongly with the conventional $\bar q q$
mesons.  However, they can also have exotic quantum numbers inaccessible
to $\bar q q$ mesons.  Confirmation
of such states would give information on the role of ``dynamical'' color in
low energy QCD.
In the quenched approximation we present a lattice calculation of the masses of
mesons with exotic quantum numbers.  These hybrid mesons can mix with four quark
($\bar q \bar q q q$) states.  The quenched approximation partially
suppresses this mixing.  Nonetheless, our hybrid interpolating fields
also couple to four quark states.
Using a four quark source operator, we demonstrate this mixing
for the $1^{-+}$ meson.  Using the conventional Wilson quark action,
we calculate both at reasonably light quark masses, intending to extrapolate to
small quark mass, and near the charmed quark mass, where we calculate
the masses of some $\bar c c g$ hybrid mesons.  The hybrid meson masses
are large --- over 4 GeV for charmonium and more than twice the
vector meson mass at our smallest quark mass, which is near the
strange quark mass.
\end{titlepage}

	
\section{Introduction}
While there is a long history of glueball mass calculations in lattice
QCD, including attempts to use lattice calculations to identify
experimentally observed mesons with glueballs\cite{WEINGARTEN,UKQCDGLUE},
hybrid mesons
have received much less attention.  These bound states of quarks and
gluons have been treated in a variety of approximations to QCD,
including the bag model, flux-tube model, and QCD sum rules\cite{kn:barnes95}.
As with glueball candidates, a hybrid characterization of an observed
state is more convincing if there is not only a match in mass, but
also in decay branching ratios.  Since the occurrence of hybrid states
is obviously a nonperturbative phenomenon, and since lattice gauge
theory provides an ab initio nonperturbative formulation of QCD, in
principle lattice gauge theory is the ideal method for calculating
their properties.  One hopes that lattice methods will eventually
provide reliable masses and branching ratios, as well as providing a
basis for testing the various approximations to QCD.

Although in principle the lattice approach is ideal, in practice there
are some difficulties.  On the lattices used here, the hybrid masses
are large compared to the lattice spacing, 
so lattice spacing errors are a serious problem.
Also, the propagators are noisy, although perhaps not so much as
glueball propagators.  This means that we do not have long plateaus
in the effective mass, and we must extract mass estimates from rather
short distances.  While we will end by making the best mass estimates
we can, at this stage we are still exploring methods and the dependence
of the results on parameters such as the lattice spacing and quark mass.

The earliest lattice calculations of hybrid mesons used static quarks,
where hybrid states appear as excitations of the gluonic
string\cite{HEAVY_HYBRIDS}.  Also in an early study, the UKQCD
collaboration studied hybrid states in the Upsilon system, in a simulation of
nonrelativistic QCD\cite{UKQCDOLD}.
More recently, the UKQCD group has presented results in the
quenched approximation for quark masses about equal to the strange
quark mass\cite{UKQCD1,UKQCD2} and we have presented preliminary results
using Wilson valence quarks and Kogut-Susskind sea quarks\cite{LAT96}.

\psfigure 5.0in 0.5in {QUARKLINEFIG}  {quarklinefig.ps} {
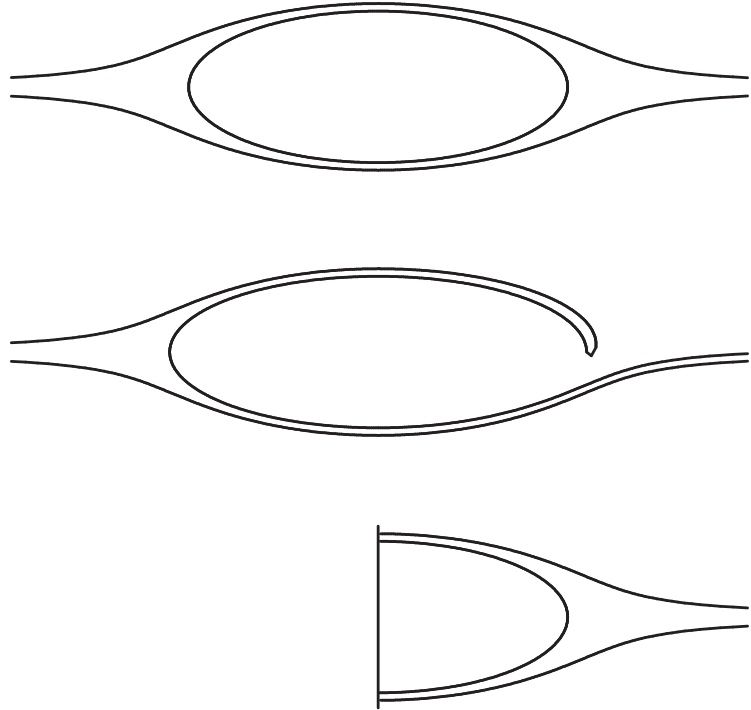
Quark line diagrams showing mixing with sea quarks (top),
``hairpin'' diagrams mixing hybrid and $\bar q \bar q q q$ states
in the quenched approximation (center), and an off diagonal propagator
with a four-quark source and a hybrid sink (bottom).  The vertical
line indicates two meson operators at the same Euclidean time but different
spatial coordinates.
}

Hybrid mesons can have the same quantum numbers as conventional
$\bar q q$ mesons, and would be expected to mix strongly with them.
(This mixing was demonstrated in Ref.~\cite{LAT96}.)  In addition,
hybrid mesons can have exotic quantum numbers.  Flavor nonsinglet
hybrids with exotic quantum numbers are especially interesting because
they cannot mix either with ordinary mesons or with glueballs.
They can, however, mix with four quark ($\bar q \bar q q q$) states.
In the quenched approximation mixing of hybrids and four quark states
through sea quark loops (Fig.~\ref{QUARKLINEFIG}a) is not present.
However, the hybrid interpolating operators may still couple to four
quark states through ``hairpin'' diagrams
(Fig.~\ref{QUARKLINEFIG}b).  This coupling can be investigated by
using four quark source operators (Fig.~\ref{QUARKLINEFIG}c),
and we find that it can even be a useful tool in computing the
mass of the exotic mesons.

Because of the partial suppression of mixing with four quark
states and because we have a supply of quenched lattices at
several lattice spacings, we have calculated exotic meson propagators
in the quenched approximation at $6/g^2=5.85$ and $6.15$.
We have done this at a set of quark masses greater than or about equal
to the strange quark mass, and at approximately the charmed quark mass.

\section{Hybrid operators and propagator computations}
To make an operator which creates a hybrid meson, we
combine a quark, an antiquark and the color electric or magnetic field to form
a color singlet with the desired spin, parity and charge conjugation.
We construct these operators by combining representations of the
continuum rotation group.
An alternative approach using the symmetry group of the hypercubic
lattice was presented by Mandula\cite{MANDULA}, and
is also developed in Ref.~\cite{UKQCD1}.

Our hybrid operators have the generic structure
$ \bar\psi^a \Gamma \psi^b F^{ab} $, where
$a$ and $b$ are triplet color indices, $\Gamma$ is some combination
of Dirac matrices and derivatives, and $F$ is the color electric
or magnetic field, a color octet.
Because we do not include ``quark-line disconnected'' diagrams in our
propagator, all our meson propagators are flavor nonsinglets.

The color electric and magnetic fields have $J^{PC} = 1^{--}$ and
$1^{+-}$ respectively.
The spin, parity and charge conjugation from
the quark and antiquark are those of the available quark bilinears,
listed here along with the corresponding mesons.
The $0^{+-}$ bilinear, $\psibar \gamma_0
\psi$, does not correspond to a $\bar q q$ state.
Instead, it is the charge
corresponding to a conserved current, the baryon number.
Therefore we expect
$\int d^3 x \, \psibar \gamma_0\psi \, | 0 \rangle = 0$.
We may, however, calculate the propagator for this exotic operator, or
use this bilinear as part of our toolkit for constructing hybrid operators.
\vskip 0.05in

\begin{center}\begin{tabular}{lll}
$0^{++}$ & ($\pbp$) & ($a_0$)\\ 
$0^{+-}$ & ($\psibar \gamma_0 \psi$) & ($J_B$) \\
$0^{-+}$ & ($\psibar \gamma_5 \psi$ , $\psibar \gamma_5\gamma_0 \psi$) & ($\pi$) \\
$1^{++}$ & ($\psibar \gamma_5\gamma_i \psi$) & ($a_1$) \\
$1^{+-}$ & ($\psibar \gamma_5\gamma_0\gamma_i \psi$) & ($b_1$) \\
$1^{--}$ & ($\psibar \gamma_i \psi$ , $\psibar \gamma_i\gamma_0 \psi$) & ($\rho$) \\
\end{tabular}\end{center}
\vskip0.05in
We can also give the quark and antiquark a relative orbital angular
momentum.
This may be useful because in the nonrelativistic quark model the $a_1$
($1^{++}$),
and hence the $0^{+-}$ and $0^{--}$ hybrids constructed below,
is a P~wave state.
The operator $\diffpartial_i = \rightpartial_i - \leftpartial_i$
inserted in the quark bilinear
brings in quantum numbers $1^{--}$,
where the negative charge conjugation
comes because C interchanges the quark and antiquark.
Thus, a P-wave operator with $a_1$ quantum numbers, $1^{++}$, is
$ \epsilon_{ijk} \psibar \gamma_j \diffpartial_k \psi $.
This operator may be advantageous because it couples the ``large''
components of the quark spinor to the large components of the antiquark
spinor.

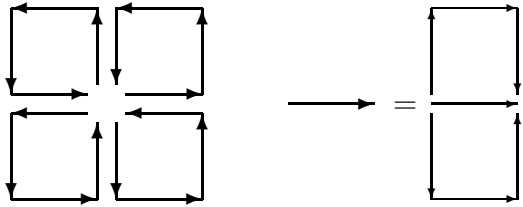
\begin{figure}[htb]
\setlength{\unitlength}{0.5in}
\begin{picture}(6.0,2.3)(-3.0,-1.15)	

\thicklines
\put(0.2,0.1){\vector(1,0){0.8}}
\put(1.0,0.1){\vector(0,1){0.9}}
\put(1.0,1.0){\vector(-1,0){0.9}}
\put(0.1,1.0){\vector(0,-1){0.8}}

\put(-0.1,0.2){\vector(0,1){0.8}}
\put(-0.1,1.0){\vector(-1,0){0.9}}
\put(-1.0,1.0){\vector(0,-1){0.9}}
\put(-1.0,0.1){\vector(1,0){0.8}}

\put(-0.2,-0.1){\vector(-1,0){0.8}}
\put(-1.0,-0.1){\vector(0,-1){0.9}}
\put(-1.0,-1.0){\vector(1,0){0.9}}
\put(-0.1,-1.0){\vector(0,1){0.8}}

\put(0.1,-0.2){\vector(0,-1){0.8}}
\put(0.1,-1.0){\vector(1,0){0.9}}
\put(1.0,-1.0){\vector(0,1){0.9}}
\put(1.0,-0.1){\vector(-1,0){0.8}}

\put(1.9,0.0){\vector(1,0){0.9}}
\put(3.0,-0.1){=}
\thinlines
\put(3.4,0.0){\vector(1,0){0.9}}
\put(3.4,0.1){\vector(0,1){0.9}}
\put(3.4,1.0){\vector(1,0){0.9}}
\put(4.3,1.0){\vector(0,-1){0.9}}
\put(3.4,-0.1){\vector(0,-1){0.9}}
\put(3.4,-1.0){\vector(1,0){0.9}}
\put(4.3,-1.0){\vector(0,1){0.9}}

\end{picture}
\caption{
``pointlike'' construction of $F_{\mu\nu}$. Each open loop represents
the product of the links, minus the adjoint of the product.
Each of these links may actually be a ``smeared'' link, as illustrated
on the right side of the figure.
}
\label{FIELDFIG}
\end{figure}

For $F_{\mu\nu}$ we use a ``pointlike'' construction, illustrated in
Fig.~\ref{FIELDFIG}.
Each open loop represents
the product of the links, minus the adjoint of the product.
To improve the overlap of the operator with the
hybrid meson, we can replace each link
by a ``smeared'' link, as illustrated on the right side of the figure.
The smeared link is the sum of the single link plus the three link
paths displaced in the spatial directions, so there are four such
staples for a link in a spatial direction and six for a link in
the time direction.  We have experimented with including staples
displaced in the time direction in the smearing, and found that
this distorts the propagators at short distances, and we will be
including short distances in our fits.  In our earlier work on
two-flavor lattices we found that iterating the smearing twice gave a
slightly better signal than a single smearing, but that four iterations
of the smearing was worse.  Therefore we have smeared twice in this
work.  We do expect that as the physical lattice spacing is
decreased more iterations of smearing would be advantageous.

\renewcommand{\arraystretch}{1.5}
\begin{table}
\begin{center}\begin{tabular}{|llll|}
\hline
Name & $J^{PC}$ (particle) & Mnemonic & Operator \\
\hline
$\pi$   &  $0^{-+}$ ($\pi$) &  $\bar q q$ pion &
    $\bar\psi^a \gamma_5 \psi^a$ \\
$\rho$   &  $1^{--}$ ($\rho$) &  $\bar q q$ rho &
    $\bar\psi^a \gamma_i \psi^a$ \\
$a_1$   &  $1^{++}$ ($a_1$) &  $\bar q q$ $a_1$ &
   $\bar\psi^a \gamma_5 \gamma_i \psi^a$ \\
$a_1(P)$   &  $1^{++}$ ($a_1$) &  P-wave $a_1$ &
   $ \epsilon_{ijk} \psibar^a \gamma_j \diffpartial_k \psi^a $ \\
\hline
$0^{-+}_H$ & $0^{-+}$ ($\pi$) & $\rho \otimes  B$ & 
   $\epsilon_{ijk} \psibar^a \gamma_i \psi^b F_{jk}^{ab}$ \\
$1^{--}_H$ & $1^{--}$ ($\rho$) & $\pi \otimes  B$ &
   $\epsilon_{ijk} \psibar^a \gamma_5 \psi^b F_{jk}^{ab}$ \\
$1^{++}_H$ & $1^{++}$ ($a_1$) & $\rho \otimes  E$ &
   $\epsilon_{ijk} \psibar^a \gamma_j \psi^b F_{0k}^{ab}$ \\
\hline
$0^{+-}_S$ & $0^{+-}$ (exotic) & $a_1 \otimes  B$ &
   $\psibar^a \gamma_5\gamma_i \psi^b \epsilon_{ijk}F_{jk}^{ab}$ \\
$0^{+-}_P$ & $0^{+-}$ (exotic) & $a_1$(P) $\otimes  B$ &
   $\psibar^b \gamma_j \diffpartial_k \psi^a F_{jk}^{ab}$ \\
$0^{+-}_B$ & $0^{+-}$ (exotic) & $J_B$ &
   $\psibar^a \gamma_0  \psi^a$ \\
\hline
$0^{--}_S$ & $0^{--}$ (exotic) & $a_1 \otimes  E$ &
   $\psibar^a \gamma_5\gamma_i \psi^b F_{i0}^{ab}$ \\
$0^{--}_P$ & $0^{--}$ (exotic) & $a_1$(P) $\otimes  E$ &
   $\epsilon_{ijk} \psibar^b \gamma_j \diffpartial_k \psi^a F_{0i}^{ab}$ \\
\hline
$\ONEMP$ & $1^{-+}$ (exotic) & $\rho \otimes  B$ &
   $\psibar^a \gamma_j\psi^b F_{ji}^{ab}$ \\
$\ONEMPTWO$ & $1^{-+}$ (exotic) & $J_B \otimes  E$ &
   $\psibar^a \gamma_0 \psi^b F_{0i}^{ab}$ \\
$\QQQQ$ & $1^{-+}$ (exotic) & $\pi \otimes a_1$ &
   $\psibar_\alpha^a(\vec x) \gamma_5 \psi_\beta^a(\vec x)
	\psibar_\beta^b(\vec y) \gamma_5 \gamma_i \psi_\lambda^b(\vec y)$ \\
\hline
\end{tabular}\end{center}
\caption{
Source and sink operators used for our propagators.
The first column lists the name used for the operator in the text, and
the second column lists the angular momentum, parity and charge conjugation.
The third column is a shorthand indicating how the operator is constructed.  In
particular, the hybrid operators are constructed from one of the quark
bilinear operators from the top block, combined with either the color
electric or color magnetic field.  The last column lists the actual operator.
In this column, $a$ and $b$ are color indices, $i$, $j$ and $k$ Cartesian
indices, and $\alpha$, $\beta$ and $\lambda$ are flavor indices included to
indicate how the propagators are connected.
}
\label{OPTABLE}
\end{table}
\renewcommand{\arraystretch}{1.0}

Table~\ref{OPTABLE} shows the various source and sink operators
we have used.  To construct a meson propagator, we first fixed
the gauge to the lattice Coulomb gauge.  We then constructed a wall
source on one time slice, with a $1$ at each lattice point for
one color and spin component.  A quark propagator was constructed by
computing $M^{-1}$ times this source.  Then the wall source was
multiplied by the source operator, which involved multiplication by
Dirac matrices and components of the field strength.  The
result of this was used as a source for an inversion to compute the
antiquark propagator.  (This involves an extra $\gamma_5$ at each end of
the propagator, from the standard identity $M^\dagger = \gamma_5 M^*
\gamma_5$.)  Finally, at each lattice point the antiquark propagator was
multiplied by the desired sink operator, and dotted with the quark
propagator.  The result was summed over each time slice to get the zero
momentum mesons.

The first three operators in Table~\ref{OPTABLE} are standard operators
for the $0^{-+}$, $1^{--}$ and $1^{++}$ $\bar q q$ mesons.  We will call
these the ``$\pi$'', ``$\rho$'' and ``$a_1$'' respectively. However, our
valence quarks are really much heavier than the physical $u$ and $d$
quarks, so they might be better thought of as $\bar s s$ or, in one
case, $\bar c c$ mesons.
The fourth operator is a P-wave operator for the $a_1$.  In the
nonrelativistic quark model the $a_1$ is a P-wave state, so the
operator with the spatial derivative will connect large components of
the Dirac spinor to large components, and may give a better signal.
This argument also applies to the hybrid operators that we will
construct using this quark bilinear as a building block.

The next set of operators are hybrid operators with the same quantum
numbers as $\bar q q$ operators.  In our previous work we verified, by
computing propagators with a hybrid operator at one end and a $\bar q q$
operator at the other, that these hybrids mix with the corresponding
$\bar q q$ operators (and don't mix with other quantum numbers)\cite{LAT96}.
The ``mnemonic'' column of the table indicates how the operator is
constructed.  For example, the first hybrid operator, which has pion
quantum numbers, can be considered as a quark and antiquark in a
$1^{--}$, or $\rho$ state (but a color octet) combined with a color
magnetic field, which has $J^{PC}=1^{+-}$, to make a $J=0$ color singlet
object.

The remaining sections of Table~\ref{OPTABLE} contain the operators with
exotic quantum numbers.  We have experimented with three $0^{+-}$
operators.  The first two are formed from the $a_1$ quark bilinear and
the color magnetic field, while the third is the $J_B$ bilinear.
Of these, the $0^{+-}_P$ source and sink gave the best signal.
(The $0^{+-}_P$ operator in the form needed for the lattice computation
is shown in the appendix.)

There are two $0^{--}$ operators, using the ``pointlike'' and the
``P-wave'' $a_1$ bilinear respectively, and the color electric field.
Again, we found that the P-wave operator gave a better signal.

Finally, there are three $1^{-+}$ operators.  The first is a quark and
antiquark in a ``$\rho$'' state, with the color magnetic field.  As is
well known\cite{HORNANDMANDULA}, a $1^{-+}$ hybrid can be constructed with the
quark and antiquark in a relative S-wave state, and this is one of
several arguments leading us to expect that it will be the lightest
exotic hybrid.  The second $1^{-+}$ operator is the charge bilinear $\bar \psi
\gamma_0 \psi$ combined with the color electric field.  Of these two
operators, we find that the first gives a better signal.
The last operator is a four-quark operator.  To use this operator as a
source, we begin with the usual wall source, which is used as
the source for a quark propagator computation.  We then multiply
the wall source by the
pion operator (multiply by $\gamma_5$), and use this as the source for
a computation of an antiquark propagator.  The resulting propagator
on the source time slice is then multiplied by the $a_1$ operator,
(multiply by $\gamma_5 \gamma_i$),
and the result is used as the source for another
antiquark propagator.
Obviously this is expensive, since it involves an extra propagator
computation.  We have not used this operator as a ``sink'' operator,
since to do this we would need a separate extra inversion at each
time slice, instead of just at the source time slice.
However, the ``crossed'' propagator with the four quark operator as
a source and the hybrid $1^{-+}$ operator as a sink turns out to
be useful.

\section{Simulation results}


We used quenched lattices with the standard plaquette gauge
action on $20^3 \times 48$ lattices at $6/g^2=5.85$ and
on $32^3 \times 64$ lattices at $\gbeta=6.15$.
These lattices were
generated for quenched spectrum studies\cite{MILCLATTICES}.

At $6/g^2=6.15$ we evaluated propagators at five values of
the Wilson hopping parameter, with the largest one chosen at
approximately the charm quark mass.   We used $\pi$, $\rho$,
$a_1$, and all of the exotic operators in Table~\ref{OPTABLE}
as source operators.  For each source, we used all of the sink
operators with the same quantum numbers, except for the $\QQQQ$
operator.
For each lattice, we used four source time slices.
For the lightest quark mass exotic propagators, we used 30 lattices,
with fewer lattices for the smaller $\kappa$ values.
Because we do not need as many lattices to get good
values for the $\pi$ and $\rho$ masses, and because we did not
implement the $\QQQQ$ source until the project was already started,
we do not have all the propagators on all of the lattices.

At $6/g^2=5.85$ we evaluated propagators with $\pi$, $\rho$,
$a_1$(P), $1^{-+}$ and $\QQQQ$ sources.
Since our previous work on $6/g^2=5.6$ two flavor lattices and
our concurrent work at $6/g^2=6.15$ had found the $1^{-+}$ to
be the lightest of the exotics and the one for which we had the
best signals, we did not do the $0^{+-}$ and $0^{--}$ propagators
on the $\gbeta=5.85$ lattices.
We used 23 lattices, with four source time slices on each lattice.

Since propagators with different values of $\kappa$ were computed
on the same quenched lattices, the mass estimates are strongly
correlated.  Also, there is the possibility of correlations among
the propagators with different source time slices on the same lattice,
which we have ignored in computing the covariance matrix.
On the other hand, the different lattices are 
uncorrelated, unlike the situation in most full QCD calculations.
To account for these correlations, especially when extrapolating
or interpolating masses to different $\kappa$ values, a jackknife
analysis is useful.
At $\gbeta=5.85$ all of the propagators were run on the same set of
lattices.
There we used a single elimination jackknife, in which
mass fits to the selected distance range and the extrapolation to
$\kc$ were repeated, each time omitting one lattice.  For the mass
estimates at $\gbeta=5.85$ in Table~\ref{QQFITS} the first
parenthesized error estimate is from the covariance matrix, and the second
is from the jackknife analysis.  These
two estimates agree well.
At $\gbeta=6.15$ we do not have all the propagators or all the kappa
values on every lattice.  Therefore we divided the lattices up into five
jackknife blocks, with each block containing about the same number of
each kind of propagator, and ran the fits for the selected distance
ranges, each time omitting one fifth of the lattices.  Again, when there
is a second parenthesized error estimate in Table~\ref{QQFITS} it is
from the jackknife analysis.

We first need an estimate of the lattice spacing.  This can be
done by extrapolating the $\rho$ mass to the physical quark mass,
or essentially to $\kappa_c$, or from the 1S-1P mass splitting
for heavy quarks.  Table~\ref{QQFITS} contains mass estimates
for the pseudoscalar, vector and $1++$ $\bar q q$ (``$\pi$'',
``$\rho$'' and ``$a_1$'') states.  These mass estimates are crude by
today's standards, but we need them only for approximately determining
the lattice spacing.  For $\gbeta=5.85$ we include $\pi$ and $\rho$ mass
estimates at additional $\kappa$ values, done on a $12^3$ spatial
lattice, coming from our $f_B$ calculations\cite{FBLATTICES}.

At $\gbeta=5.85$ we estimate $1/\kc=6.205(3)$ from extrapolating
the squared pion masses from the largest four $\kappa$ values.
If we estimate the lattice spacing by extrapolating the $\rho$ mass to
$\kc$, we find $a^{-1}=1.80(6)$ GeV or 2.00(11) GeV, depending on
whether $m_\rho$ or $m_\rho^2$ is extrapolated linearly in $1/\kappa$.
These extrapolations are plotted in Fig.~\ref{MASSES585}.

At $\gbeta=6.15$ we do not have the luxury of extra pion masses near
$\kappa_c$.  A linear fit of $m_\pi^2$ in $1/\kappa$ does not work
well, so we fit the pion mass at the four largest $\kappa$'s to a
quadratic in $1/\kappa$ to estimate $1/\kappa_c = 6.4895(10)$.
This fit had a $\chi^2$ of 0.5 with one degree of freedom.
An extrapolation of $m_\rho$ to this $\kappa_c$ gives
$a m_\rho(\kappa_c) = 0.274(4)$, or $a^{-1}=2.81(4)$ GeV,
where the error is a jackknife estimate.
These extrapolations are plotted in Fig.~\ref{MASSES615}.
Another possibility, following the Fermilab group, is to use the
splitting between the average S~wave charmonium masses and the P~wave
masses to estimate the lattice spacing.  At $\kappa=0.1350$, using the
$1^{++}$ meson as the P~wave mass, and $\frac{3}{4} m_{1^{--}} + \frac{1}{4} m_{0^{-+}}$ as the S~wave mass, with 457 MeV
as the experimental value for charmonium, we get $a^{-1} = 2.85(1)$ GeV.

\psfigure 5.0in 0.5in {MASSES585}  {masses_b585.ps} {
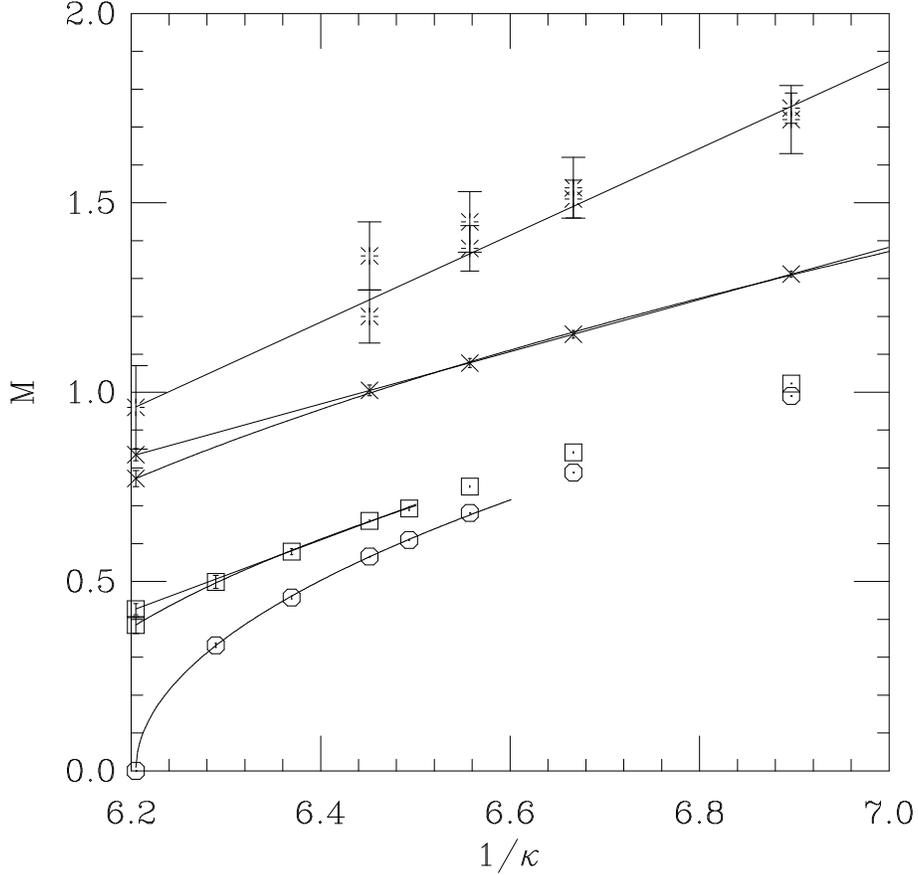
Particle masses at $\gbeta=5.85$, and extrapolations to $\kappa_c^{-1}$. 
The octagons are pion masses, and the line is a fit to $m_\pi^2$
linear in $1/\kappa$.
Squares and crosses are $\rho$ and $a_1$ masses respectively.  For the
$\rho$ and $a_1$ we show two extrapolations of the mass to $\kappa_c$,
one
with $m_{\rho,a_1}$ linear in $1/\kappa$ and the other with
$m_{\rho,a_1}^2$ linear in $1/\kappa$.
Finally, the bursts are the $\ONEMP$ exotic meson.
We show two fits of the $\ONEMP$ meson for each $\kappa$.
The lower one uses  the two-source, two-mass fits as illustrated
in Fig.~\protect\ref{EMASSFIG1350}, while the upper one uses
only the $\ONEMP\rightarrow\ONEMP$ propagator.
}
\psfigure 5.0in 0.5in {MASSES615}  {masses_b615.ps} {
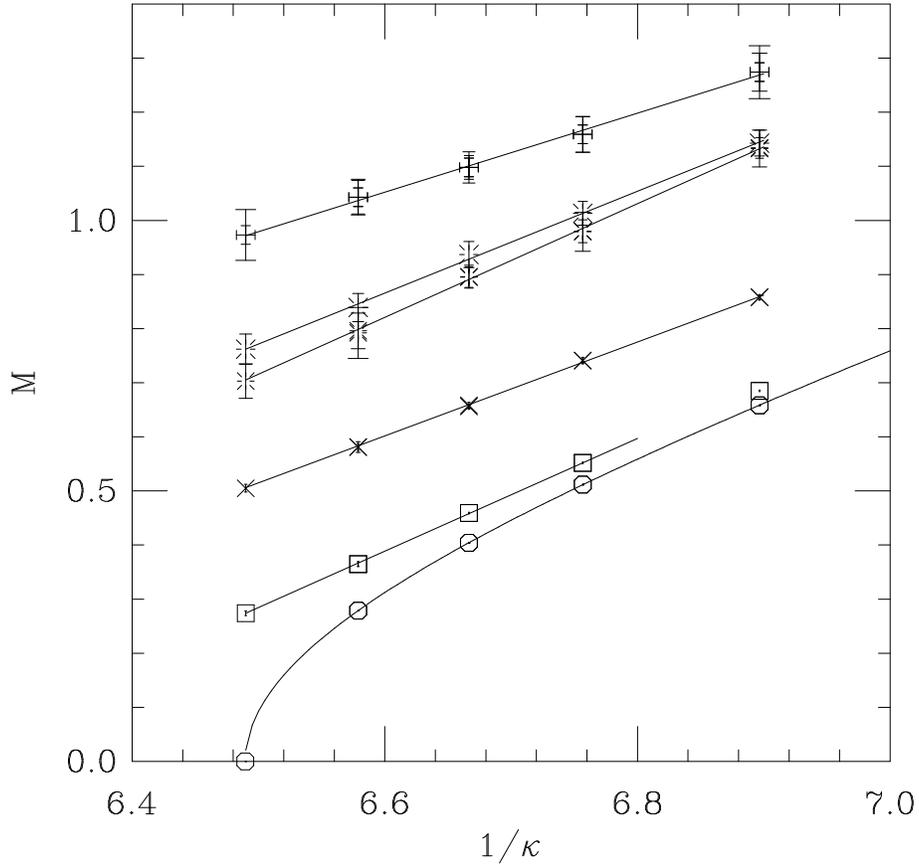
Particle masses at $\gbeta=6.15$, and extrapolations to $\kappa_c^{-1}$. 
The symbols are the same as in Fig.~\protect\ref{MASSES585},
and the fancy plusses are a $\ZEROPM$ exotic.
Again we show
two fits of the $\ONEMP$ meson for each $\kappa$, the lower one
fitting both source operators simultaneously and the upper one fitting
only the $\ONEMP\rightarrow\ONEMP$ propagator.
The $\kappa=0.1350$, or $1/\kappa=7.407$, charmonium point is not
in the range shown here.
}

\psfigure 5.0in 0.5in {PROPFIG1350}  {props.615.k1350.ps} {
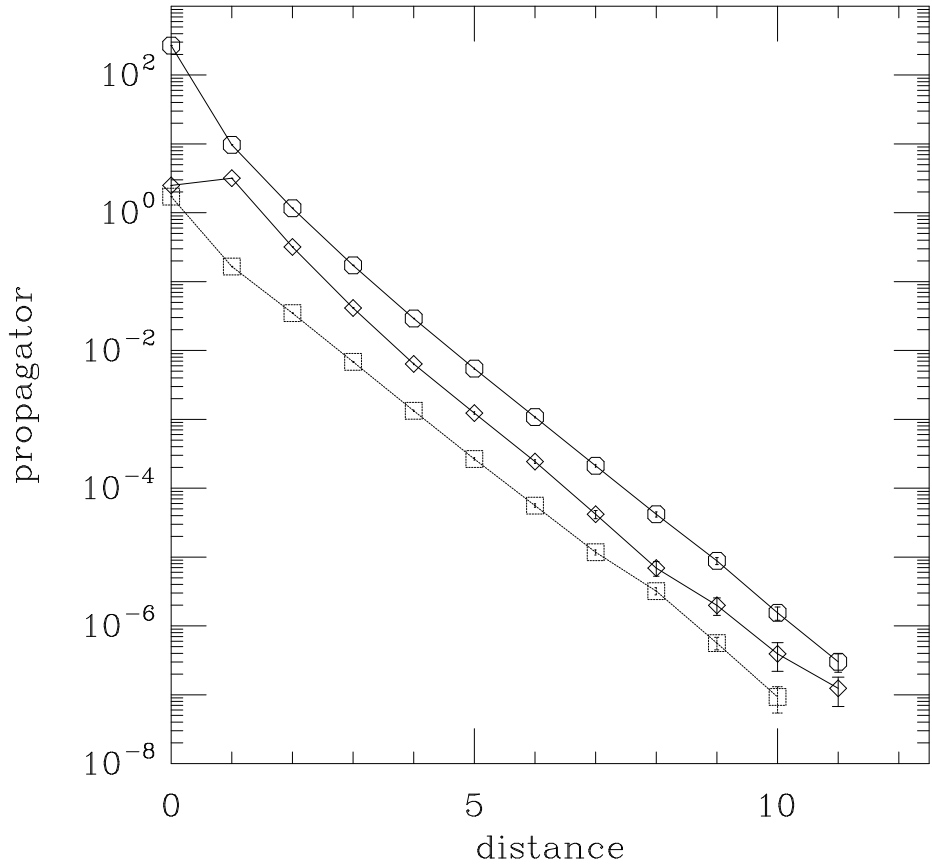
Propagators for the $1^{-+}$ exotic meson at $\gbeta=6.15$ and
$\kappa=0.1350$.  The octagons are for a $\ONEMP$ source and sink, the
diamonds for a $\ONEMP$ source with a $\ONEMPTWO$ sink, and the squares for a
$\QQQQ$ source with a $\ONEMP$ sink.
}

\psfigure 5.0in 0.5in {PROPFIG1520}  {props.615.k1520.ps} {
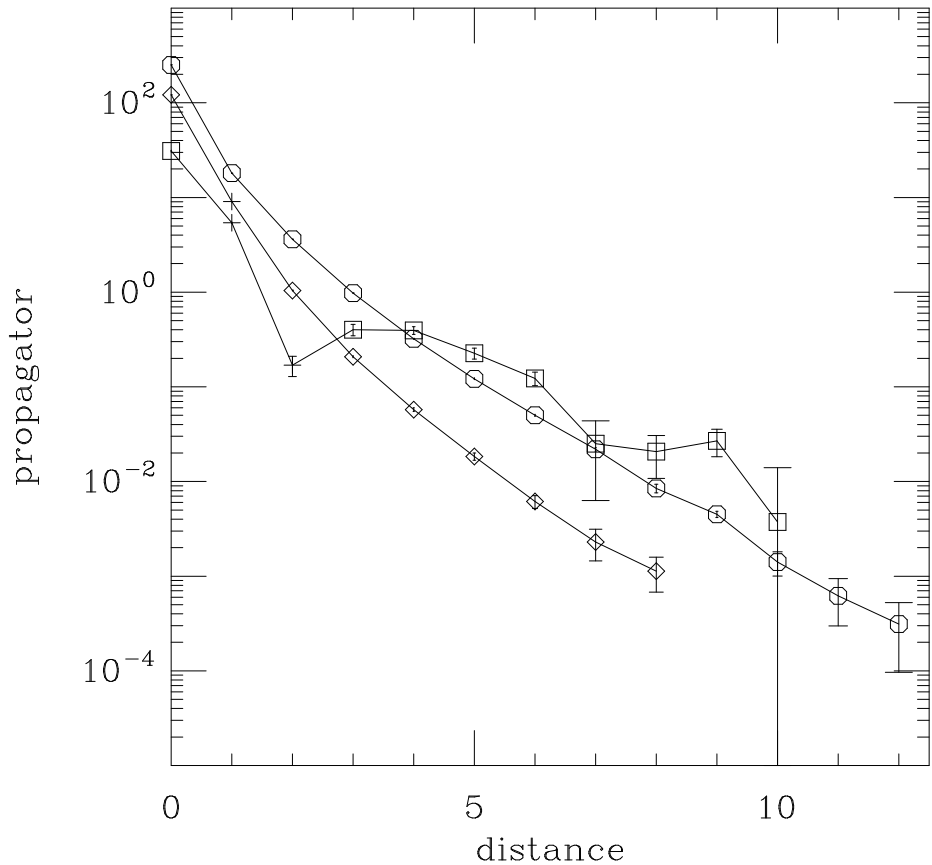
Propagators for the $1^{-+}$ exotic meson at $\gbeta=6.15$ and
$\kappa=0.1520$.
The octagons are for a $\ONEMP$ source and sink, the
diamonds for a $\ONEMPTWO$ source and sink, and the squares for a
$\QQQQ$ source with a $\ONEMP$ sink.
Plusses indicate the absolute value of a 
propagator that changes sign with increasing distance.
}

\psfigure 5.0in 0.5in {EMASSFIG1350}  {emass.1mp.1350.ps} {
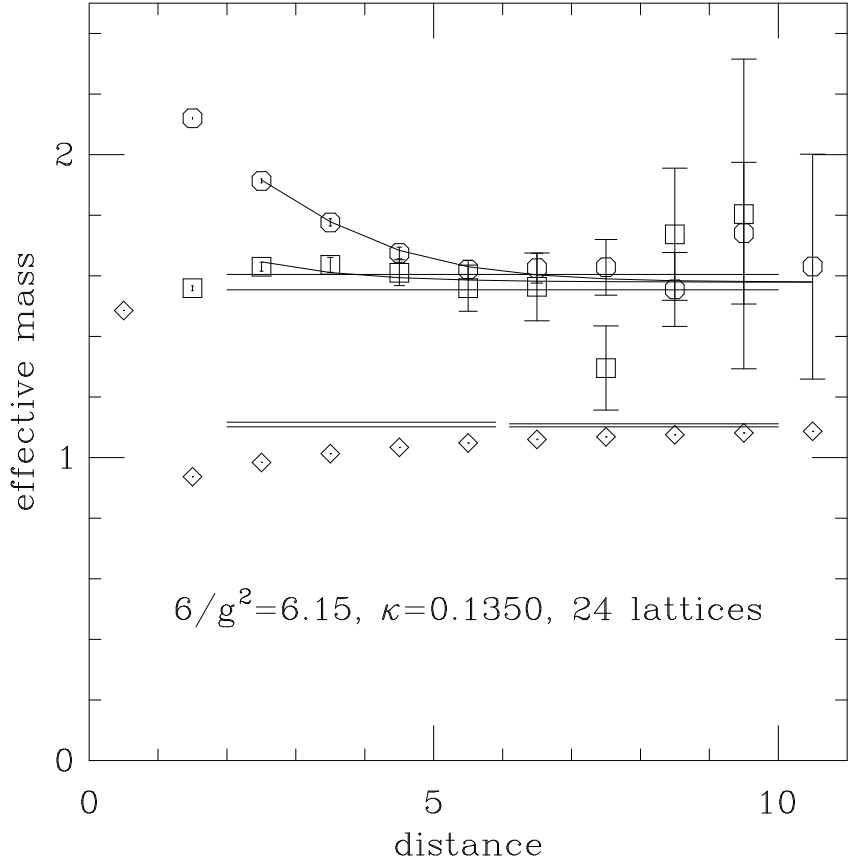
Effective masses for the $\rho$ and $\ONEMP$ exotic at $\gbeta=6.15$
and $\kappa=0.1350$,
approximately the charm quark mass.
The octagons are the effective mass from propagators with the $\ONEMP$
operator as both source and sink, and the squares are the effective mass
with the $\QQQQ$ source and $\ONEMP$ sink.  These propagators were
fit with two source amplitudes and two masses as described in the text,
over the distance range 2 to 11.
The solid lines near the plot symbols are the effective masses
reconstructed from the fit.  The upper pair of horizontal lines indicates the
$\pm 1 \sigma$ range for the ground state mass in this fit.
The diamonds are the $\rho$ (more accurately, the $\psi$)
effective mass.  The horizontal bars near the diamonds are the
$\pm 1 \sigma$ limits on the ground state mass
from fits to the $\rho$ propagator.  The lines
running from $d=2$ to 6 are a two mass fit to the propagator over this
range, while the closely spaced lines for $d>6$ are from a single
exponential fit over distance range 21 to 25.
}
\psfigure 5.0in 0.5in {EMASSFIG1520}  {emass.1mp.1520.ps} {
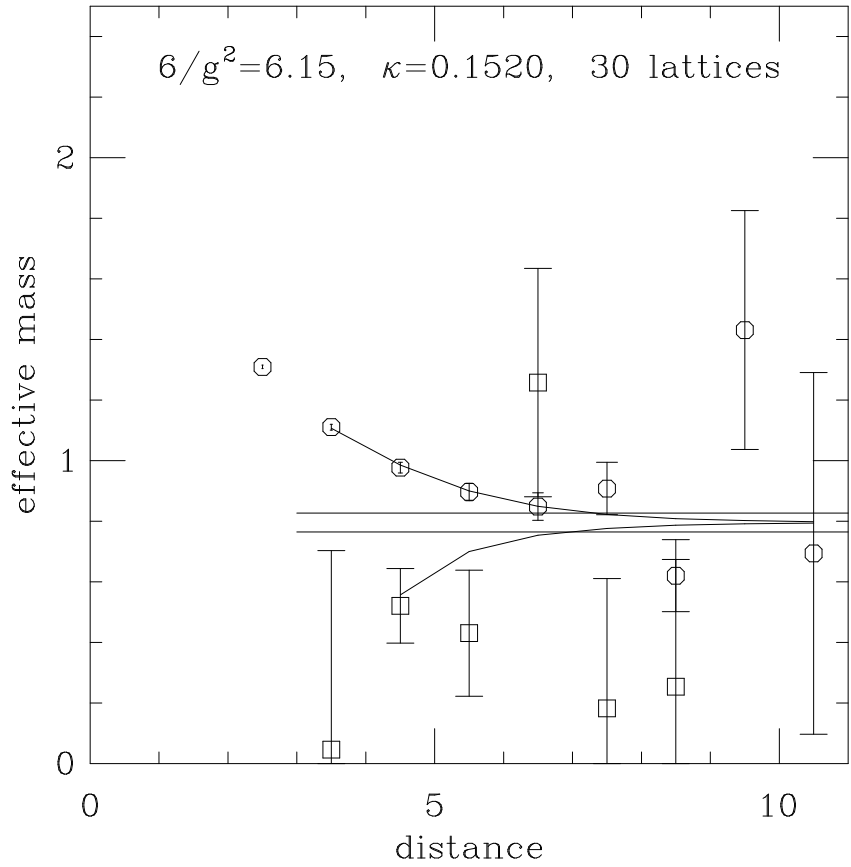
Effective masses for the $1^{-+}$ meson at $\gbeta=6.15$
and $\kappa=0.1520$, approximately the strange quark mass.
The symbols and lines have the same meaning as in
Fig.~\protect\ref{EMASSFIG1350}.  These effective masses 
correspond to the propagators in Fig.~\protect\ref{PROPFIG1520}.
}
\psfigure 5.0in 0.5in {EMASSFIG5851450}  {emass.585.1450.ps} {
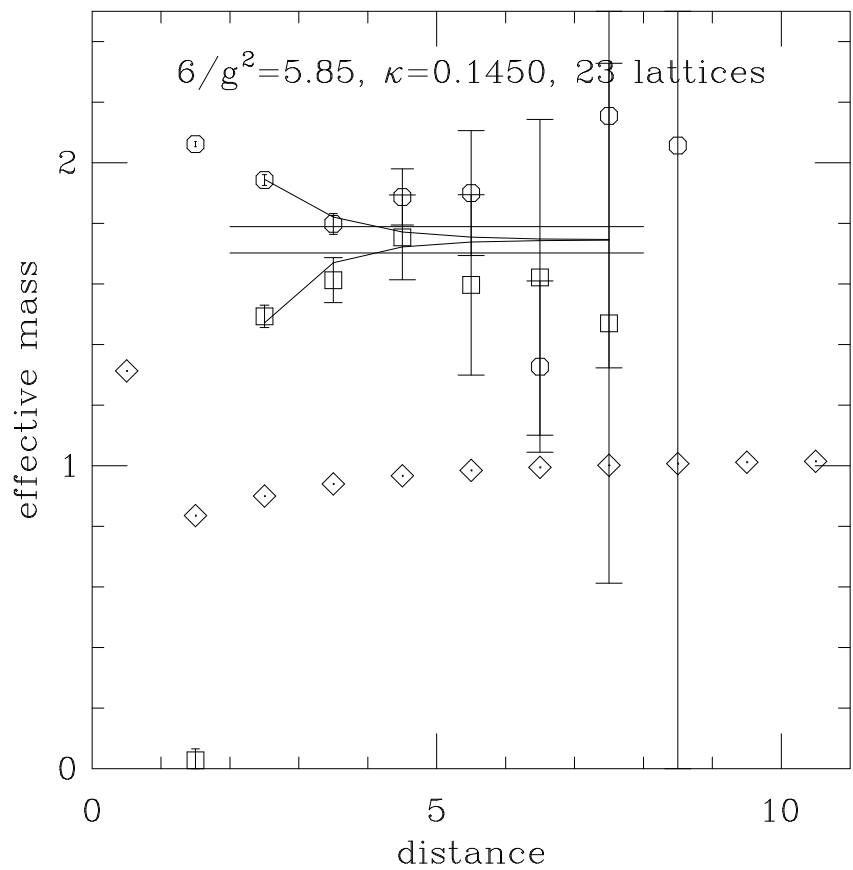
Effective masses for the $1^{-+}$ meson and the $\rho$ at $\gbeta=5.85$
and $\kappa=0.1450$.
The symbols and lines have the same meaning as in
Fig.~\protect\ref{EMASSFIG1350}.
}
\psfigure 5.0in 0.5in {EMASSFIG5851550}  {emass.585.1550.ps} {
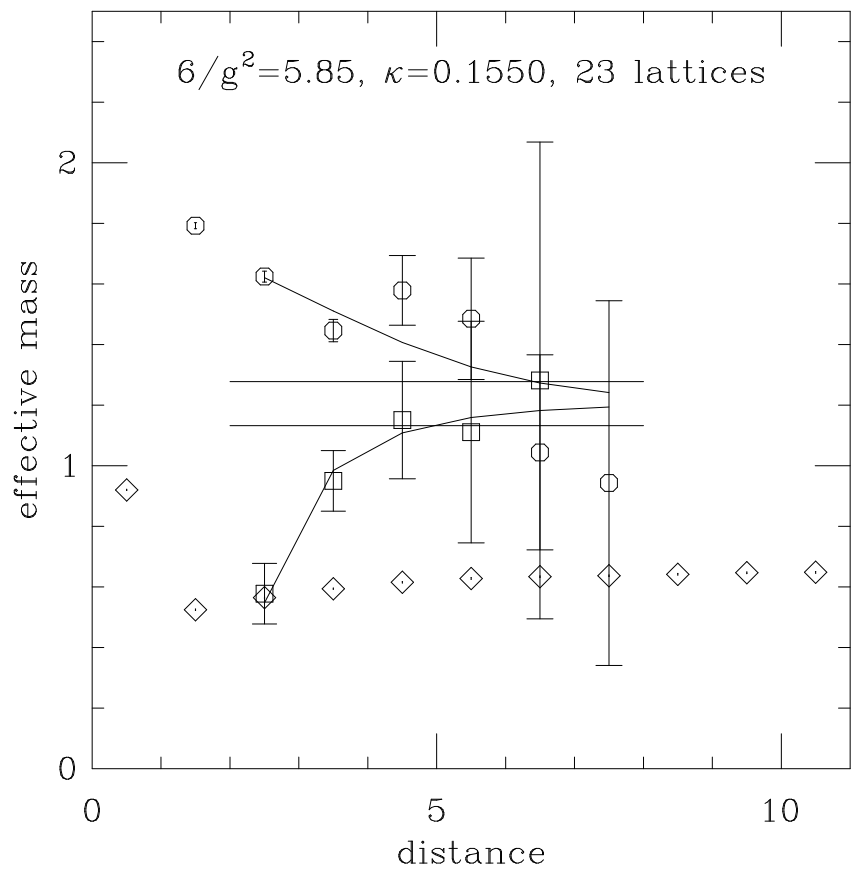
Effective masses for the $1^{-+}$ meson and the $\rho$ at $\gbeta=5.85$
and $\kappa=0.1550$.
The symbols and lines have the same meaning as in
Fig.~\protect\ref{EMASSFIG1350}.
}

Fig.~\ref{PROPFIG1350} shows exotic propagators at $\gbeta=6.15$ and
$\kappa=0.1350$.  For the $1^{-+}$ exotic we show three propagators.
One has the $\ONEMP$ operator as its source and sink, the second the
$\ONEMP$ operator as the source and the $\ONEMPTWO$ operator as the sink,
and the third has the $\QQQQ$ operator as its source
and the $\ONEMP$ operator as its sink.
Fig.~\ref{PROPFIG1520} is a similar figure for our smallest
quark mass, $\kappa=0.1520$, except that instead of the
$\ONEMP\rightarrow\ONEMPTWO$ propagator we have a
$\ONEMPTWO\rightarrow\ONEMPTWO$ propagator.
Compared to the propagators for conventional mesons, these
exotic propagators are quite noisy, and we must use fairly short
distances for the mass fits.
As seen in Figs.~\ref{PROPFIG1350} and \ref{PROPFIG1520},
this problem becomes worse as the quark mass is made lighter.
In fact, these exotic propagators fall
below their statistical error at a distance smaller than the minimum
distance we use for a single mass fit for the $\pi$ or $\rho$.
This means that contamination by excited states might be a serious problem.
We have therefore done fits to two exponentials for the exotic
states.  We can do this only over a small range of minimum
distance.  In particular, if we take the minimum distance too
large we get fits with a very large excited state mass, which
is essentially just a delta function removing the shortest distance
point from the fit, and giving the same result as a one mass fit
with a minimum distance one unit larger.  Of course, we reject
such fits since they are basically one mass fits.

\begin{table}
\begin{center}\begin{tabular}{|lllcll|}
\hline
$6/g^2$ & particle & $\kappa$ & Fit Range & $\chi^2/{\rm DOF}$ & Mass \\
\hline
6.15	& $\pi$ & 0.1350	& 21--25 & 1.7/3 & 1.0967(45)(56) \\
	&	& 0.1450	& 18--24 & 1.5/5 & 0.6583(7)(8) \\
	&	& 0.1480	& 18--24 & 0.3/5 & 0.5117(7)(7) \\
	&	& 0.1500	& 18--24 & 1.4/5 & 0.4042(8)(5) \\
	&	& 0.1520	& 18--24 & 6.4/5 & 0.2788(10)(3) \\
\hline
6.15	& $\rho$ & 0.1350	& 21--25 & 2.4/3 & 1.1067(5)(8) \\
	&	& 0.1450	& 18--24 & 4.4/5 & 0.6850(10)(15) \\
	&	& 0.1480	& 18--24 & 2.2/5 & 0.5518(13)(17) \\
	&	& 0.1500	& 18--24 & 2.4/5 & 0.4591(17)(14) \\
	&	& 0.1520	& 18--24 & 2.1/5 & 0.3644(34)(48) \\
	&	& 0.154		& extrap. & --- & 0.2738(-)(42) \\
\hline
6.15	& $a_1$	& 0.1350	& 16--20 & 2.6/3 & 1.274(8)(10) \\
	&	& 0.1450	& 8--14 & 9.0/5 & 0.858(5)(5) \\
	&	& 0.1480	& 8--14 & 8.5/5 & 0.741(6)(5) \\
	&	& 0.1500	& 8--14 & 7.4/5 & 0.658(6)(5) \\
	&	& 0.1520	& 8--14 & 4.1/5 & 0.581(10)(6) \\
	&	& 0.154		& extrap. & --- & 0.505(-)(7) \\
\hline
5.85	& $\pi$	& 0.1450	& 13--18 & 1.0/4 & 0.990(1)(1) \\
	&	& 0.1500	& 13--18 & 1.6/4 & 0.788(1)(1) \\
	&	& 0.1525	& 13--18 & 1.8/4 & 0.681(1)(1) \\
	&	& $0.1540^*$	& 5-18	& 17.4/24 & 0.610(2) \\
	&	& 0.1550	& 13--18 & 2.4/4 & 0.566(1)(1) \\
	&	& $0.1570^*$	& 5-18	& 17.9/24 & 0.458(4) \\
	&	& $0.1590^*$	& 5-18	& 15.3/24 & 0.332(6) \\
\hline
5.85	& $\rho$ & 0.1450	& 13--18 & 2.8/4 & 1.023(1)(2) \\
	&	& 0.1500	& 13--18 & 2.2/4 & 0.841(2)(2) \\
	&	& 0.1525	& 13--18 & 2.1/4 & 0.751(2)(2) \\
	&	& $0.1540^*$	& 8-19	& 5.4/10 & 0.693(5) \\
	&	& 0.1550	& 13--18 & 2.3/4 & 0.660(3)(3) \\
	&	& $0.1570^*$	& 8-19	& 5.6/10 & 0.579(8) \\
	&	& $0.1590^*$	& 8-19	& 3.7/10 & 0.499(17) \\
\hline
5.85	& $a_1$	& 0.1450	& 6--11 & 1.7/4 & 1.312(8)(8) \\
	&	& 0.1500	& 6--11 & 3.0/4 & 1.153(10)(9) \\
	&	& 0.1525	& 6--11 & 4.5/4 & 1.077(12)(10) \\
	&	& 0.1550	& 6--11 & 5.5/4 & 1.005(14)(13) \\
\hline
\end{tabular}\end{center}
\caption{
Mass estimates for ordinary $\bar q q$ mesons.
The $a_1$ mass estimates used the $a_1$(P) source and sink.
An asterisk indicates a point from a $12^3$ lattice.
Where present, a second parenthesized error is a jackknife estimate.
Fits at $\kappa=0.154$ for $\gbeta=6.15$
are jackknife extrapolations to $\kc$. 
}
\label{QQFITS}
\end{table}

\begin{table}
\begin{center}\begin{tabular}{|lrcclll|}
\hline
 $\kappa$ & Source(s)$\rightarrow$Sink & masses & Fit Range & $\chi^2/{\rm DOF}$ & Mass
& M* \\
\hline
0.1450 & $\ONEMP \rightarrow \ONEMP$ & 1
	    & 3--9 & 3.9/5 & 1.81(3) & \\
	& $\ONEMP \rightarrow \ONEMP$ & 1
	    & 4--10 & 3.5/5 & 1.88(8) & \\
	& $\QQQQ \rightarrow \ONEMP$ & 1
	    & 3--7 & 0.7/3 & 1.65(5) & \\
	& $\ONEMP \rightarrow \ONEMP$ & 2
	    & 1--9 & 7.3/5 & 1.72(9) & 2.45(19) \\
	& $\QQQQ,\ONEMP \rightarrow \ONEMP$ & 2
	    & 1--8 & 9.1/10 & 1.71(3) & 2.42(6) \\
	& $\QQQQ,\ONEMP \rightarrow \ONEMP$ & 2
	    & 2--8 & 6.6/8 & 1.75(4)(5) & 2.86(44) \\
\hline
0.1500	& $\ONEMP \rightarrow \ONEMP$ & 2
	    & 1--9 & 5.8/5 & 1.54(8) & 2.33(19) \\
	& $\QQQQ,\ONEMP \rightarrow \ONEMP$ & 2
	    & 1--8 & 7.4/10 & 1.47(3) & 2.20(4) \\
	& $\QQQQ,\ONEMP \rightarrow \ONEMP$ & 2
	    & 2--8 & 6.2/8 & 1.51(5)(6) & 2.31(21) \\
\hline
0.1525	& $\ONEMP \rightarrow \ONEMP$ & 2
	    & 1--9 & 5.2/5 & 1.45(8) & 2.28(18) \\
	& $\QQQQ,\ONEMP \rightarrow \ONEMP$ & 2
	    & 1--8 & 7.4/10 & 1.33(3) & 2.10(4) \\
	& $\QQQQ,\ONEMP \rightarrow \ONEMP$ & 2
	    & 2--8 & 6.0/8 & 1.38(6)(7) & 2.08(15) \\
\hline
0.1550	& $\ONEMP \rightarrow \ONEMP$ & 2
	    & 1--9 & 5.0/5 & 1.36(9) & 2.22(17) \\
	& $\QQQQ,\ONEMP \rightarrow \ONEMP$ & 2
	    & 1--8 & 9.3/10 & 1.19(4) & 2.00(3) \\
	& $\QQQQ,\ONEMP \rightarrow \ONEMP$ & 2
	    & 2--8 & 6.0/8 & 1.20(7)(8) & 1.88(10) \\
\hline
\end{tabular}\end{center}
\caption{
Mass estimates for the exotic $1^{-+}$ meson for $\gbeta=5.85$.
Where two source operators are listed, a simultaneous fit was
done to propagators from both sources, with the masses forced to
be the same for each source.  Where two masses were used in the fit,
the last column shows the excited state mass produced by the fit.
Where present, a second parenthesized error is a jackknife estimate.
}
\label{ONEMPFITS585}
\end{table}

\begin{table}
\begin{center}\begin{tabular}{|lrcclll|}
\hline
 $\kappa$ & Source(s)$\rightarrow$Sink & masses & Fit Range & $\chi^2/{\rm DOF}$ & Mass
& M* \\
\hline
0.1350	& $\ONEMP \rightarrow \ONEMP$ & 2
	    & 1--11 & 2.3/7 & 1.61(2) & 2.52(4) \\
	& $\ONEMP \rightarrow \ONEMP$ & 2
	    & 2--11 & 1.3/6 & 1.58(3) & 2.42(11) \\
	& $\ONEMP \rightarrow \ONEMP$ & 2
	    & 3--11 & 1.2/5 & 1.59(4) & 2.54(42) \\
	& $\QQQQ,\ONEMP \rightarrow \ONEMP$ & 2
	    & 1--11 & 18.1/16 & 1.63(1) & 2.57(3) \\
	& $\QQQQ,\ONEMP \rightarrow \ONEMP$ & 2
	    & 2--11 & 6.8/14 & 1.58(3) & 2.38(9) \\
	& $\QQQQ,\ONEMP \rightarrow \ONEMP$ & 2
	    & 3--11 & 6.5/12 & 1.57(4) & 2.28(23) \\
\hline
0.1450	& $\ONEMP \rightarrow \ONEMP$ & 2
	    & 1--11 & 11.5/7 & 1.189(16) & 2.19(4) \\
	& $\ONEMP \rightarrow \ONEMP$ & 2
	    & 2--11 & 4.5/6 & 1.143(24) & 2.24(7) \\
	& $\QQQQ,\ONEMP \rightarrow \ONEMP$ & 2
	    & 2--11 & 15.2/14 & 1.109(16) & 1.90(4) \\
	& $\QQQQ,\ONEMP \rightarrow \ONEMP$ & 2
	    & 3--11 & 8.8/12 & 1.134(19)(34) & 2.04(18) \\
	& $\QQQQ,\ONEMP \rightarrow \ONEMP$ & 2
	    & 4--11 & 4.2/10 & 1.161(39) & 1.62(47) \\
\hline
0.1480	& $\ONEMP \rightarrow \ONEMP$ & 2
	    & 2--11 & 4.8/6 & 1.013(22) & 2.16(6) \\
	& $\ONEMP \rightarrow \ONEMP$ & 2
	    & 3--11 & 3.0/5 & 1.039(25) & 2.30(43) \\
	& $\QQQQ,\ONEMP \rightarrow \ONEMP$ & 2
	    & 2--11 & 30.4/14 & 0.955(15) & 1.76(4) \\
	& $\QQQQ,\ONEMP \rightarrow \ONEMP$ & 2
	    & 3--11 & 15.1/12 & 0.980(21)(36) & 1.74(12) \\
	& $\QQQQ,\ONEMP \rightarrow \ONEMP$ & 2
	    & 4--11 & 6.3/10 & 1.019(29) & 1.67(33) \\
\hline
0.1500	& $\ONEMP \rightarrow \ONEMP$ & 2
	    & 2--11 & 8.8/6 & 0.937(24) & 2.16(6) \\
	& $\ONEMP \rightarrow \ONEMP$ & 2
	    & 3--11 & 8.7/5 & 0.943(29) & 1.91(24) \\
	& $\QQQQ,\ONEMP \rightarrow \ONEMP$ & 2
	    & 2--11 & 48.9/14 & 0.868(17) & 1.70(03) \\
	& $\QQQQ,\ONEMP \rightarrow \ONEMP$ & 2
	    & 3--11 & 22.7/12 & 0.897(22)(19) & 1.63(10) \\
	& $\QQQQ,\ONEMP \rightarrow \ONEMP$ & 2
	    & 4--11 & 15.5/10 & 0.943(26) & 1.88(42) \\
\hline
0.1520	& $\ONEMP \rightarrow \ONEMP$ & 2
	    & 2--11 & 6.9/6 & 0.839(26) & 2.13(6) \\
	& $\ONEMP \rightarrow \ONEMP$ & 2
	    & 3--11 & 6.4/5 & 0.820(37) & 1.63(16) \\
	& $\QQQQ,\ONEMP \rightarrow \ONEMP$ & 2
	    & 2--11 & 38.7/14 & 0.789(23) & 1.66(4) \\
	& $\QQQQ,\ONEMP \rightarrow \ONEMP$ & 2
	    & 3--11 & 16.0/12 & 0.796(33)(47) & 1.54(11) \\
	& $\QQQQ,\ONEMP \rightarrow \ONEMP$ & 2
	    & 4--11 & 14.3/10 & 0.827(40) & 1.76(50) \\
\hline
0.154	& $\QQQQ,\ONEMP \rightarrow \ONEMP$ & 2
	    & extrap. & --- & 0.705(na)(32) & \\
\hline
\end{tabular}\end{center}
\caption{
Mass estimates for the exotic $1^{-+}$ meson for $\gbeta=6.15$.
The format is the same as Table~\protect\ref{ONEMPFITS585}.
The final line is an extrapolation to $\kappa_c^{-1}$, using the
distance range 3--11 for the four largest $\kappa$ values.
}
\label{ONEMPFITS615}
\end{table}

\begin{table}
\begin{center}\begin{tabular}{|lrcclll|}
\hline
 $\kappa$ & Source(s)$\rightarrow$Sink & masses & Fit Range & $\chi^2/{\rm DOF}$ & Mass
& M* \\
\hline
0.1350	& $0^{+-}_P \rightarrow 0^{+-}_P$ & 1
	    & 5--9 & 3.6/3 & 1.72(2) & \\
	& $0^{+-}_P \rightarrow 0^{+-}_P$ & 1
	    & 6--10 & 2.7/3 & 1.69(3) & \\
	& $0^{+-}_P \rightarrow 0^{+-}_P$ & 1
	    & 7--11 & 1.2/3 & 1.63(5) & \\
	& $0^{+-}_P \rightarrow 0^{+-}_P$ & 2
	    & 2--11 & 4.5/6 & 1.66(3)(5) & 2.52(6) \\
\hline
0.1450	& $0^{+-}_P \rightarrow 0^{+-}_P$ & 1
	    & 5--9 & 2.9/3 & 1.32(3) & \\
	& $0^{+-}_P \rightarrow 0^{+-}_P$ & 1
	    & 6--10 & 3.6/3 & 1.30(4) & \\
	& $0^{+-}_P \rightarrow 0^{+-}_P$ & 1
	    & 7--11 & 2.7/3 & 1.23(7) & \\
	& $0^{+-}_P \rightarrow 0^{+-}_P$ & 2
	    & 2--11 & 3.9/6 & 1.27(4)(5) & 2.24(7) \\
\hline
0.1480	& $0^{+-}_P \rightarrow 0^{+-}_P$ & 1
	    & 5--9 & 1.9/3 & 1.20(2) & \\
	& $0^{+-}_P \rightarrow 0^{+-}_P$ & 1
	    & 6--10 & 2.2/3 & 1.18(4) & \\
	& $0^{+-}_P \rightarrow 0^{+-}_P$ & 1
	    & 7--11 & 1.9/3 & 1.11(8) & \\
	& $0^{+-}_P \rightarrow 0^{+-}_P$ & 2
	    & 2--11 & 3.5/6 & 1.16(3)(3) & 2.16(6) \\
\hline
0.1500	& $0^{+-}_P \rightarrow 0^{+-}_P$ & 1
	    & 5--9 & 0.7/3 & 1.13(2) & \\
	& $0^{+-}_P \rightarrow 0^{+-}_P$ & 1
	    & 6--10 & 5.4/3 & 1.08(4) & \\
	& $0^{+-}_P \rightarrow 0^{+-}_P$ & 1
	    & 7--11 & 4.9/3 & 1.01(9) & \\
	& $0^{+-}_P \rightarrow 0^{+-}_P$ & 2
	    & 2--11 & 7.5/6 & 1.10(3)(2) & 2.16(6) \\
\hline
0.1520	& $0^{+-}_P \rightarrow 0^{+-}_P$ & 1
	    & 5--9 & 2.7/3 & 1.08(3) & \\
	& $0^{+-}_P \rightarrow 0^{+-}_P$ & 1
	    & 6--10 & 1.9/3 & 0.98(6) & \\
	& $0^{+-}_P \rightarrow 0^{+-}_P$ & 1
	    & 7--11 & 1.7/3 & 0.94(13) & \\
	& $0^{+-}_P \rightarrow 0^{+-}_P$ & 2
	    & 2--11 & 4.2/6 & 1.04(3)(3) & 2.13(7) \\
\hline
\end{tabular}\end{center}
\caption{
Mass estimates for the exotic $0^{+-}$ meson for $\gbeta=6.15$.
The format is the same as Table~\protect\ref{ONEMPFITS585}.
}
\label{ZEROPMFITS615}
\end{table}

In these fits, we would like to use
information from the different source and sink operators, by
simultaneously fitting to two or more combinations of operators with
different amplitudes for each source and sink but the same masses for
all of them.
\BE
\langle {\cal O}_i(0) {\cal O}_j(t) \rangle =
  A^0_i  A^0_j e^ { -m_0 t }
+  A^1_i  A^1_j e^ { -m_1 t } \ldots \ \ \ ,\EE
where $i$ and $j$ label the source and sink operators respectively, and
$m_0$ and $m_1$ are the ground state and excited state masses.
For this to be useful, the relative overlaps of the
different operators with the ground state and excited state should
be as different as possible.  For the $1^{-+}$ propagators it turns out
that the $\ONEMP$ and $\ONEMPTWO$ operators have essentially the same
effective masses.  Since the propagators with $\ONEMPTWO$ operators are
noisier than those with $\ONEMP$ source and sink, including these
propagators in the fitting did not help (any gain in statistics was
not worth the extra degrees of freedom in the fitting.)
However, the correlator generated from the $\QQQQ$ source,
which we introduced to investigate coupling to four quark states,
does have an effective mass at short distance significantly different from
that generated from the hybrid $\ONEMP$ source.
This can be seen in
Fig.~\ref{EMASSFIG1350}, which shows the effective masses corresponding to
the $\ONEMP\rightarrow\ONEMP$ and $\QQQQ\rightarrow\ONEMP$ propagators
at our heaviest quark mass, $\kappa=0.1350$, and in
Fig.~\ref{EMASSFIG1520} which shows the effective masses at our
lightest quark mass, $\kappa=0.1520$, corresponding to
the propagators in Fig.~\ref{PROPFIG1520}.
As might be expected from this, simultaneously fitting the $\ONEMP$ and
$\QQQQ$ source propagators, each with the $\ONEMP$ sink operator,
to two masses gave the best mass estimates.
The ground state mass from this fit,
and the effective masses from the fit, are also shown
in Figs.~\ref{EMASSFIG1350} and \ref{EMASSFIG1520}.
Of course, this two source fit makes the assumption that the excited state
(or combined effect of many excited states treated as a single state
in the fitting program) is the same in both propagators.  Therefore
we tabulate results both from the simultaneous fits to two source operators
and from fits using only the $\ONEMP$ source operator.
In the cases of the $0^{+-}$ and $0^{--}$ propagators, we have not
investigated four quark source operators.  For the $0^{+-}$ and $0^{--}$
hybrid operators, the P~wave source
and sink operators generally gave the best statistical errors, and
so were the only ones we fit.

One might still worry that we are not extracting the correct ground
state masses from such short distances.  As a partial check,
we take the $\rho$ propagator, for which a quite convincing plateau
in the effective mass is seen at larger distances, and make a two mass
fit to this propagator at the same distances we use for the hybrid
fits.  (Since the hybrid fits are dominated by the more accurate points
at the smaller distances in the fitting range, we fit the rho mass to
the smaller distances in the range.)  In this case, at $\gbeta=6.15$ and
$\kappa=0.1350$, which is the charm quark mass,
we find that the $\rho$ (more accurately, the ${\rm J}/\psi$)
mass from a single exponential
fit to distance range 21 to 25 is 1.1067(5)  ($\chi^2/d=2.4/3$), while
a two mass fit over distance range 2 to 6 gives a ground state
mass 1.109(8) with an excited state mass of 1.264(21) ($\chi^2/d=1.5/1$),
in excellent agreement with the single exponential fit from long distances.
The $\rho$ (${\rm J}/\psi$) effective mass and these fits are also shown in
Fig.~\ref{EMASSFIG1350}.
While this result is encouraging, we
should caution the reader that the number of excited states and the mass
gap between the ground and excited states might be very different for
the $\rho$ meson and the exotic mesons.

Another important test is to verify that the mass estimates are
independent of the fitting range used.
Here we are not in as good a position, since we
generally have only two or three minimum distances where we can get a two mass
fit with reasonable $\chi^2$.  However, within the fairly poor
statistical errors, the exotic mass estimates are generally consistent
among these fits.

The $\gbeta=5.85$ fits to the $\ONEMP$ are done in similar fashion.
In Figs.~\ref{EMASSFIG5851450} and \ref{EMASSFIG5851550}
we show the effective masses and the fits to the propagators for
our largest and smallest $\kappa$ values, $\kappa=0.1450$ and $0.1550$,
at $\gbeta=5.85$.

Tables~ \ref{ONEMPFITS585} and \ref{ONEMPFITS615} contain selected
mass fits for the $1^{-+}$ mesons at $\gbeta=5.85$ and $6.15$ respectively.
When two masses were used in the fit, both the ground state and
the excited state mass are tabulated.  However, this excited state
mass is almost certainly some sort of weighted average of many
states, and should not be taken seriously as a mass estimate.

Finally, Table~\ref{ZEROPMFITS615} contains selected fits for
the $\ZEROPM$ exotic at $\gbeta=6.15$  This particle is clearly
heavier than the $\ONEMP$, and our estimates for its mass are
worse.  This is partly because we have only the one source operator
for this meson, and partly because (nonrelativistically) the quark
and antiquark are in a relative P~wave state, and we were led
to use a more complicated source operator.

We were unable to get credible mass estimates from the $\ZEROMM$
propagators, suggesting that this state, if it exists at all, is
even heavier than the $\ONEMP$ and $\ZEROPM$.


\section{Discussion and Conclusions}

\psfigure 5.0in 0.5in {RATIOS3}  {ratios3.ps} {
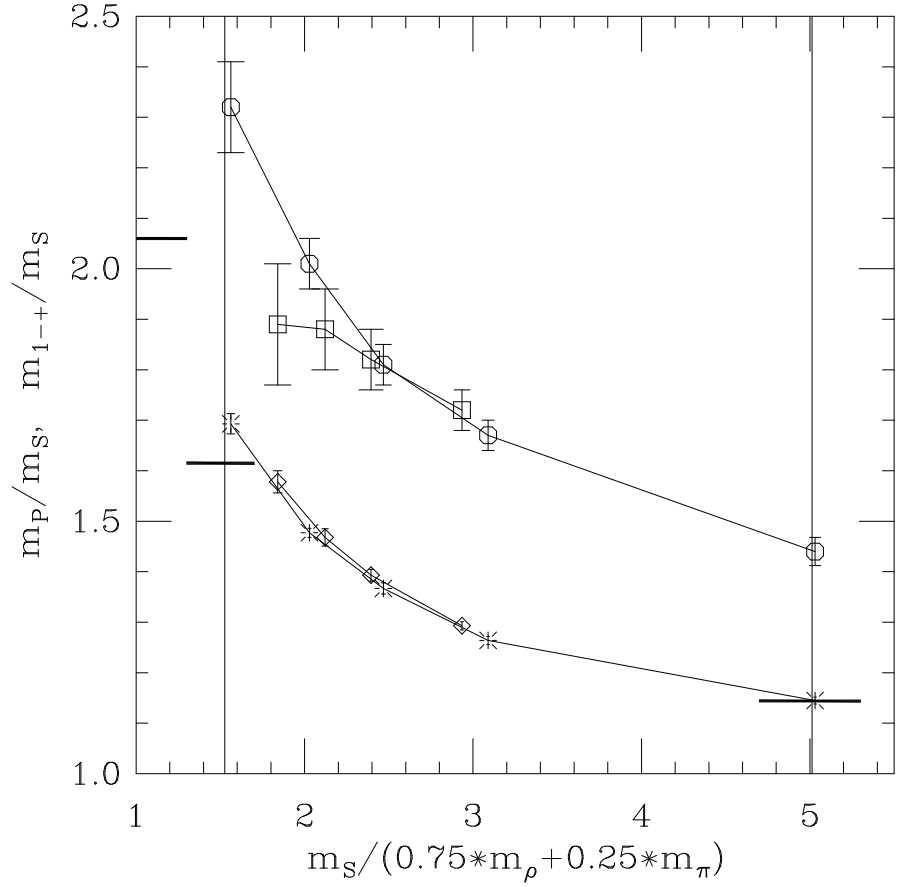
Ratios of the $1^{++}$ (P~wave) meson mass and the $\ONEMP$ exotic
mass to the average S~wave meson mass.
Diamonds and bursts are the $1^{++}$ at $\gbeta=5.85$ and
$6.15$ respectively, and squares and octagons are the $\ONEMP$ at
$5.85$ and $6.15$ respectively.  The horizontal and vertical scales and
the vertical and horizontal lines are described in the text.
}

In this concluding section we discuss the conversion of our mass
estimates from lattice units to physical mass units, estimate
systematic errors, and describe briefly the observational status
of $\ONEMP$ exotic hadrons.

Figures~\ref{MASSES585} and \ref{MASSES615} collect and display our results
for nonexotic mesons and for the $\ONEMP$ and the $\ZEROPM$ exotic mesons.
We also show extrapolations to $\kappa_c$, where the error on the
extrapolations comes from the jackknife analysis.
For the $\ONEMP$ hybrid we plot fits to both the one source operator
and two source operator mass estimates.  The difference between these
fits is an indication of the possible systematic error from excited
states in the propagators.  While this difference is small for
the charmonium point, at the strange quark mass for $\gbeta=6.15$ it
amounts to 117 MeV, and when extrapolated to $\kappa_c$ becomes 165 MeV.

One of the known problems with the Wilson
quark action is that it consistently underestimates spin splittings
of hadrons\cite{EL_KHADRA}.
This suggests that the average mass of the S~wave
mesons might be a better mass standard than the $\rho$ mass alone.
(We have already used this logic in section 3 when we used the splitting between
the P-wave and the average S-wave charmonium mass as a length scale.)
In Fig.~\ref{RATIOS3} we plot the ratio of the $1^{++}$ P-wave meson ($a_1$)
to the average S-wave meson mass,
$\frac{3}{4} m_{1^{--}} + \frac{1}{4} m_{0^{-+}}$.
For the horizontal scale we use the average S-wave mass divided
by the average S-wave mass at the physical point, where $m_\pi/m_\rho=0.18$.
In other words, the units of the horizontal axis are 
$\frac{3}{4} m_\rho + \frac{1}{4} m_\pi = 610$ MeV.
The vertical scale is the ratio of the P-wave $1^{++}$ meson mass
or the $1^{-+}$ exotic meson mass to the S-wave mass.
The vertical lines indicate the strange quark and charmonium points.
In locating the strange quark line, we have followed the UKQCD procedure
of using an ``unmixed'' $\eta_{strange}$ mass of 680 MeV\cite{UKQCD2}.
By definition, the light quark point is the left hand side of the graph.
Where they intersect the left side of the graph and the two
vertical lines, the three bold horizontal lines indicate the
experimental values of the $1^{++}$ meson mass
divided by the S-wave mass for
light quarks, strange quarks and charm quarks respectively.

A nice feature of this graph is the good agreement of the P-wave
masses between the $\gbeta=5.85$ and $6.15$ lattices.
For heavy quarks the agreement of the $1^{-+}$ exotic is equally
good, but there is some difference for the light quarks.  We suspect
that the $\gbeta=5.85$ points are incorrect here.
In Fig.~\ref{EMASSFIG5851550} we see that the $\ONEMP$ fit
at $\gbeta=5.85$ and $\kappa=0.1550$ is
questionable at best, and  Fig.~\ref{MASSES585} or Table~\ref{ONEMPFITS585}
show that had we used the one source fits for this point
we would have obtained a 13\% larger mass.
The exact agreement between our predicted and the observed
charmonium mass is undoubtedly fortuitous, but this agreement
and the trend toward agreement between the 5.85 and 6.15
estimates at larger quark mass encourages us to quote a
charmed hybrid meson mass with suitable caveats.

Using this average S-wave mass as the length scale
for the charmonium exotics gives an {\bf uncorrected} mass
of 4390(80) MeV for the $\ONEMP$ and 4610(110) MeV for the $\ZEROPM$.
We emphasize that these quoted errors are statistical only, and do not
take into account contamination from excited states in the chosen
fitting ranges, or discretization errors in the gauge and quark action,
or effects of the quenched approximation.
As mentioned above, the errors from excited state contamination, which can be
crudely estimated by looking at how the estimated mass varies
as a function of minimum distance in the fit or whether the
second source operator is included, are probably about
the same size as the statistical errors.
The close agreement of the $1^{++}$ masses and the $1^{-+}$ masses
for heavier quarks in Figure~\ref{RATIOS3} suggests that errors from
the nonzero lattice spacing might be small.  However, we also note that
in our studies of charmed pseudoscalar meson decay constants we find a
discretization error in the decay constants of 10\% to 15\% at
$\gbeta=6.15$\cite{MILCFUTURE}.
It might be that decay constants are
more sensitive to lattice spacing errors than mass, since the 
decay constants are basically wave functions at the origin, and are
strongly affected by the coarseness of the lattice at short distance.
Although it is little better than a quess, we propose using
15\% of the splitting between the hybrid state and the $\bar c c$
states, or 200 MeV, as an estimate of the systematic error
from both excited state contamination and lattice artifacts.
The largest systematic error probably comes
from using the quenched approximation, and this is the hardest error to
estimate.  It will surely be large, since, as discussed in the
introduction, these hybrid states can mix with four quark states.
In this work we looked at mixing with states containing four heavy
quarks, but in the real world the important four quark states
would contain the charmed quark and antiquark and a light sea
quark and antiquark.
This charmonium $\ONEMP$ is quite far above the $D \bar D$ threshold.
Because of the remaining systematic uncertainties, it is not
clear whether it is above the S~wave + P~wave $D \bar D$ threshold,
which in many model calculations determines whether its decay width is
large\cite{DECAYS,kn:barnes95}.

At the strange quark mass (our largest $\kappa$ at $\gbeta=6.15$),
we estimate the mass of the $\ONEMP$ hybrid to be
$2170 \pm 80 \pm{\rm systematic}$
MeV.  As noted at the beginning of this section,
the systematic error from excited state contamination
is on the order of 100 MeV, and the error from nonzero lattice spacing
is probably as large or larger.
Considering the large errors, this estimate is consistent
with the mass quoted by the UKQCD collaboration,
2000(200) MeV\cite{UKQCD2}.

If we take seriously the extrapolation of the $\ONEMP$ mass to light
quarks in Fig.~\ref{MASSES615}, we get a mass of 1970(90) MeV for
the light quark exotic hybrid, again with large systematic errors.
Considering the lack of agreement in the $\ONEMP$ mass estimates
at $\gbeta=5.85$ and $6.15$ at light quark mass and the effects
of extrapolating in kappa, we use 300 MeV as an estimate of the
error on this number from lattice artifacts.

We briefly review the observational evidence relevant to our results.
The particle data table~\cite{kn:pdt} does not list any confirmed
$1^{-+}$ hybrid meson states. However a number of potential
candidates are mentioned.  There is some evidence for a $1^{-+}$ hybrid
state with a mass of around $1.4$ GeV, and another with a mass around
$1.9$ GeV.
The original evidence for a $1^{-+}$ state at 1.4 GeV found by the 
GAMES collaboration~\cite{kn:gams} was criticized in~\cite{kn:problem}.
However recent work by the 
E852 collaboration~\cite{kn:Eeightfivetwo} reports evidence for
a $1^{-+}$ hybrid state with a mass of $1370\pm 16_{-30}^{+15}$ MeV.  
This paper\cite{kn:Eeightfivetwo} also lists other experiments
that have reported a low mass for the $1^{-+}$ state.
Our result favors a hybrid assignment for states
around $1.9$ GeV~\cite{kn:alde89,kn:lee94}. 
However we stress that more simulations are required to quantify and reduce the 
systematic errors in our results, before definitive results for the mass of
the $1^{-+}$ state can be obtained from quenched lattice QCD.

In Table~\ref{tb:modelonemp} (obtained from~\cite{kn:barnes95} ) we
collect results for the mass of the light $1^{-+}$ hybrid obtained from
a variety of models. We note that our prediction for the mass of the 
$1^{-+}$ state is consistent with the flux tube estimate. Only 
the bag model calculation obtains a mass close to the E852 experimental 
result.

%
%
\begin{table}[thb]
\begin{center}
\begin{tabular}{ll} 
Mass (GeV) & Method \\
\hline
\multicolumn{2}{l}{ Light quark $\ONEMP$ mass}\\
$1.3 \rightarrow 1.8$   &  Bag model  \\
$1.8 \rightarrow 2.0 $   & Flux tube model   \\
$1.8 \rightarrow 1.9 $   & Flux tube model of Close \etal\cite{kn:barnes95}  \\
$2.1 \rightarrow 2.5 $   & QCD sum rules (mostly after 1984)   \\
$ 1.97(9_{stat.})(30_{lattice})(??_{quench}) $   & This work \\
\hline
\multicolumn{2}{l}{ Charmonium $\ONEMP$ mass}\\
$\approx 3.9$   &  Adiabatic bag model  \\
$4.2 \rightarrow 4.5 $   & Flux tube model   \\
$4.1 \rightarrow 4.2 $   & Flux tube model of Barnes \etal\cite{kn:barnes95}
\\
$4.19 \pm syst. $   & Heavy quark LGT\cite{HEAVY_HYBRIDS} \\
$4.1 \rightarrow 5.3 $   & QCD sum rules (mostly after 1984)   \\
$ 4.39(8_{stat.})(20_{lattice})(??_{quench}) $   & This work \\
\hline
\end{tabular}
\end{center}
\caption{Predictions for the mass of the light and charmonium
$1^{-+}$ states from various approaches to QCD,
obtained from~\protect\cite{kn:barnes95}.}
\label{tb:modelonemp}
\end{table}

Unfortunately, no good candidates for the theoretically expected
$1^{-+}$ hybrid charmonium state have been observed~\cite{kn:pdt}.  Such a
state is also expected in flux tube models, which are most plausible
for heavy quark systems and furthermore provide information about
branching ratios.
Table~\ref{tb:modelonemp} also contains some predictions
(taken from ~\cite{kn:barnes95}) for the 
mass of the $1^{-+}$ state in the charmonium system.

Although we have concentrated on exotic hybrids in this work, lattice
methods, particularly with dynamical quark loops included, could also
address interesting issues regarding nonexotic states.  For example,
Close and Page characterize the $\psi(4040)$ and $\psi(4160)$ states
as mixtures of $\bar c c$ and $\bar c c g$\cite{kn:close96a}.
Close proposes that
hybrid states could help explain the ``anomalous'' production of
charmonium observed by the CDF group\cite{kn:close95b}.  In both cases the
occurrence of a nonexotic hybrid state in the range 4 to 4.3 GeV is
essential.

Future efforts in lattice calculations of exotic hybrids should be
directed at reducing the variety of systematic errors, including
finite lattice spacing, excited state contamination, and the effects
of quenching.  Particularly important is to understand the extent of
mixing with four-quark states.
%

\section*{Acknowledgements}
This work was supported by the U.S. Department of Energy under contracts
DE-AC02-76CH-0016,
DE-AC02-86ER-40253,
DE-FG03-95ER-40906,
DE-FG05-85ER250000,\\
DE-FG05-96ER40979,
DE-2FG02-91ER-40628,
DE-FG02-91ER-40661,
and National Science Foundation grants
NSF-PHY96-01227,	
NSF-PHY97-22022.	
Computations were done on the Paragon at Oak Ridge National
Laboratory and the Paragon and T3E at the San Diego Supercomputer
Center.  We thank Maarten Golterman, Jeff Mandula and Phillip
Page for valuable
conversations. One of us (DT) is grateful for the hospitality of the 
University of Tsukuba, where this work was completed.
Two of us (CD) and (CM) are, likewise, grateful for the hospitality
of the Zentrum f\"ur Interdisziplin\"are Forschung at the University
of Bielefeld.

\section*{Appendix}

Here we write the $\ZEROPM$ P-wave operator, $0^{+-}_P$, in the form
needed for the lattice computation.  We are interested in zero spatial
momentum, so we are summing over spatial coordinates, and can freely
translate the summation variable.  We wish to write the operator
as $\bar\psi(\vec x)$ times a sum of fields at $\vec x$ and
neighboring points.  In these expressions, it is understood that
fields at neighboring points must be parallel transported to $\vec x$,
so that $\bar\psi(\vec x) \psi(\vec x + \hat k)$ means
$\bar\psi(\vec x) U_k(\vec x) \psi(\vec x + \hat k)$.

 \BEA {\cal O}_P^{+-} =
  \sum_{\vec x} \sum_{jk} &&\psibar^b \gamma_j \LP \rightpartial_k -
 \leftpartial_k \RP \psi^a F_{jk}^{ab} \EL
  = \sum_{\vec x} \sum_{jk} 
 && \psibar^b(\xvec) \gamma_j \psi^a(\xvec+\khat) F_{jk}^{ab}(\xvec) \EL
 -&& \psibar^b(\xvec) \gamma_j \psi^a(\xvec-\khat) F_{jk}^{ab}(\xvec) \EL
 -&& \psibar^b(\xvec+\khat) \gamma_j \psi^a(\xvec) F_{jk}^{ab}(\xvec) \EL
 +&& \psibar^b(\xvec-\khat) \gamma_j \psi^a(\xvec) F_{jk}^{ab}(\xvec) \EL
  = \sum_{\vec x} \sum_{jk} 
 && \psibar^b(\xvec) \gamma_j \psi^a(\xvec+\khat) F_{jk}^{ab}(\xvec) \EL
 -&& \psibar^b(\xvec) \gamma_j \psi^a(\xvec-\khat) F_{jk}^{ab}(\xvec) \EL
 -&& \psibar^b(\xvec) \gamma_j \psi^a(\xvec-\khat) F_{jk}^{ab}(\xvec-\khat) \EL
 +&& \psibar^b(\xvec) \gamma_j \psi^a(\xvec+\khat) F_{jk}^{ab}(\xvec+\khat) \EL
  = \sum_{\vec x} \sum_{jk} 
 && \psibar^b(\xvec) \gamma_j \LP  \psi^a(\xvec+\khat) \LP
 F_{jk}^{ab}(\xvec) + F_{jk}^{ab}(\xvec+\khat) \RP \RP \EL
 -&& \LP  \psi^a(\xvec-\khat) \LP
 F_{jk}^{ab}(\xvec) + F_{jk}^{ab}(\xvec-\khat) \RP \RP 
 \EEA

Similarly, for the $0^{--}$ P-wave source:
 \BEA {\cal O}_P^{--} =
  \sum_{\vec x} \sum_{ijk} \epsilon_{ijk} &&\psibar^b \gamma_j \LP \rightpartial_k -
 \leftpartial_k \RP \psi^a F_{0i}^{ab} \EL
  = \sum_{\vec x} \sum_{ijk} \epsilon_{ijk}\Big(
 && \psibar^b(\xvec) \gamma_j \psi^a(\xvec+\khat) F_{0i}^{ab}(\xvec) \EL
 -&& \psibar^b(\xvec) \gamma_j \psi^a(\xvec-\khat) F_{0i}^{ab}(\xvec) \EL
 -&& \psibar^b(\xvec+\khat) \gamma_j \psi^a(\xvec) F_{0i}^{ab}(\xvec) \EL
 +&& \psibar^b(\xvec-\khat) \gamma_j \psi^a(\xvec) F_{0i}^{ab}(\xvec) \ \Big)\EL
  = \sum_{\vec x} \sum_{ijk} \epsilon_{ijk}\Big(
 && \psibar^b(\xvec) \gamma_j \psi^a(\xvec+\khat) F_{0i}^{ab}(\xvec) \EL
 -&& \psibar^b(\xvec) \gamma_j \psi^a(\xvec-\khat) F_{0i}^{ab}(\xvec) \EL
 -&& \psibar^b(\xvec) \gamma_j \psi^a(\xvec-\khat) F_{0i}^{ab}(\xvec-\khat) \EL
 +&& \psibar^b(\xvec) \gamma_j \psi^a(\xvec+\khat) F_{0i}^{ab}(\xvec+\khat) \ \Big)\EL
  = \sum_{\vec x} \sum_{ijk} \epsilon_{ijk}\Big(
 && \psibar^b(\xvec) \gamma_j \LP  \psi^a(\xvec+\khat) \LP
 F_{0i}^{ab}(\xvec) + F_{0i}^{ab}(\xvec+\khat) \RP \RP \EL
 -&& \LP  \psi^a(\xvec-\khat) \LP
 F_{0i}^{ab}(\xvec) + F_{0i}^{ab}(\xvec-\khat) \RP \RP \ \Big) 
 \EEA

\end{document}